
\magnification 1200
\hsize 5 in
\hoffset .25 in
\baselineskip=10 pt
\centerline{{\bf Triviality Bound on Lightest Higgs Mass in NMSSM}}
\vskip 1.0cm
\centerline{S.R. Choudhury, Mamta and Sukanta Dutta}
\vskip .5cm
\centerline{{\sl Department of Physics and Astrophysics,}}
\vskip .2cm
\centerline{{\sl University of Delhi, Delhi-110007, INDIA.}}
\vskip 3cm
\centerline{{\bf Abstract}}
\vskip .5cm
\baselineskip=2\baselineskip
We study the implication of triviality on Higgs sector in
next to minimal supersymmetric model (NMSSM) using variational
field theory. It is shown that mass of the lightest Higgs boson
in NMSSM has an upper bound $\sim 10\, M_W$ which is of the same
order as that in standard model.
\vfil
\eject

\beginsection {1. Introduction}

\indent It is now widely believed that the $ \phi^4$- theory in four
space-time dimensions is trivial. If one accepts this, one is
forced to conclude that the GSW-model cannot be an exact field
theory  but  at  best a reasonable effective theory valid upto an
energy scale of $\Lambda_c$ ({\it i.e.} all integrations over
intermediate four-momenta are evaluated by putting a cut-off
$\Lambda_c$). We do not have any definite idea of the
scale of $\Lambda_c$ at persent except theoretical conjectures.
The standard model (SM) has also another mass parameter,
namely  the  mass  $M_H$  of  the  Higgs  boson,  which  can take
arbitrary values in the model. This particle has  not  yet  been
experimentally observed  but  one  can  keep  one's  faith
 in  the  SM alive by proposing that the Higgs particle lies
beyond the present experimental limits.
It  would  then seem that as far as the Higgs sector is concerned
the
experimental non-observability of the Higgs upto any energy would
never cast any shadow on the SM. Things are different if the SM
is  regarded  as  an  effective  theory  valid  upto  some  scale
$\Lambda_c$.
It would then be reasonable to demand that $M_H$ be not too close
to
$\Lambda_c$  for  the effective theory to be a reasonable one\thinspace; we
then
have a limit say $M_H<\Lambda_c/ 5$. With such a restriction, one
can establish
that the Higgs quartic  coupling cannot be arbitrarily large,
{\it i.e.}, it has
a maximum allowed value, which translates into an upper bound for
$M_H/M_W$.
This is the triviality bound studied by various non-perturbative
methods. Results are similar, namely that $M_H$ cannot be heavier
 than value
in the range 800\thinspace GeV--\thinspace 1\thinspace TeV [1].
Methods used in arriving at
these results include RGE equation [2], improved perturbative
approach [3] and recerntly
two of us have used a variational approach [4] to arrive at a
similar result. This last
approach is extremely simple and admits of easy generalisation to
situations more complicated  than SM.

Supersymmetric (SUSY) generalisation of the SM have been studied
in recent times [5].
The  most  economic  SUSY-extension of the SM is minimal (MSSM)
one [5].
In this version, the quartic couplings are
restricted  by  the gauge coupling with the result that the Higgs
cannot be arbitrarily heavy. At tree level, one has the
relationship
  $$M_H^2 \le M_Z^2\cos^2 2\beta \le M_Z^2 \,\eqno{(1)}$$
where $\tan\beta$
is  the  ratio  of  the  vacuum expectation values (VEV's) of the
neutral components  of  the  scalar  fields $H_1$ and $H_2$
that the MSSM involves.
Going   beyond   the  tree  approximation  does  not  change  (1)
qualitatively. Thus, Quiros [6] gives the bound
  $$M_H \le 125\,{\rm GeV} \eqno{(2)}$$
for $m_t=174\thinspace$GeV and a cut-off $\Lambda_c\approx 10^{19}\,$GeV.

An alternative supersymmetric model proposed is the next to
minimal supersymmetric model (NMSSM) which has two
SU(2)$\,\bigotimes\,$U(1)
Higgs doublets and one Higgs singlet [5]. The inclusion of a
Higgs singlet is suggested in many superstring models and
grand unified models. The NMSSM has more coupling parameters
than the MSSM and hence it is an intersting theoretical
question to enquire into the upper bounds of the Higgs
spectrum of the lightest of them. We  expect  this  to be much
higher than the one given in (2) and this is the subject matter
of our investigation.

  The  method  we follow here is a variational one. Starting with
Hamiltonian of the NMSSM we use a gaussian trial wave functional for
the ground state and obtain estimates of mass spectra in terms of
the bare parameters of the theory. The strategy then is to vary
bare   parameters over their entire range, impose restrictions
that the masses cannot get very close to the cut-off (say less
than $\Lambda_c/5$) and obtain the highest mass of the
lightest Higgs particle. The parameter space is however very
large, and we will be making specific choices of parameters in
the hope that our results will be typical of the model itself.

 As this investigation was in progress, a paper by Y.Y. Wu
on the triviality bound in NMSSM appeared in print [7]. The approach
used by this author is different from ours and triviality requirement
used by him was to ensure that the Higgs coupling constant remains finite
below the triviality scale. This leads to a bound on Higgs mass
much lower than the one obtained by us. However the author has used
perturbation theory which as he himself states may not be desirable.
We have used non-perturbative approach and this is the reason, apart
from the difference in the concept of triviality, for the disagreement
in results.

\beginsection {2. The Model}

\indent  In the NMSSM, the potential of the Higgs sector [7] is
 $$\eqalignno{V= &\thinspace\bigl\vert hN \bigr\vert^2
 \bigl(\Phi_1^\dagger\Phi_1
+\Phi_2^\dagger\Phi_2\bigr)+\bigl\vert h\Phi_1^\dagger\Phi_2
+\lambda N^2 \bigr\vert^2+{1\over 8} g_1^2
\bigl(\Phi_1^\dagger\Phi_1-\Phi_2^\dagger\Phi_2\bigr)^2 &\cr
&+{1\over 8}
g_2^2\Bigl[\bigl(\Phi_1^\dagger \Phi_1+\Phi_2^\dagger\Phi_2\bigr)^2
-4\bigl(\Phi_1^\dagger\Phi_2)(\Phi_2^\dagger\Phi_1\bigr)\Bigr]
+m_1^2\Phi_1^\dagger\Phi_1+m_2^2\Phi_2^\dagger\Phi_2 &\cr
&-m_3^2\bigl(\Phi_1^\dagger\Phi_2 +h.c.\bigr)+m_4^2N^{\ast}N
+m_5^2\bigl(N^2+N^{\ast2}\bigr)
&(3)\cr}$$
\vskip .2 cm
\noindent Here $\,\Phi_1=\bigl(\phi_1^\dagger ,\,\phi_1^0\bigr)$  and
$\Phi_2=\bigl(\phi_2^\dagger,\,\phi_2^0\bigr)\,$
are two SU(2)$\,\bigotimes\,$U(1) doublets,
N a complex singlet, $m$'s are mass parameters, $g$'s gauge couplings
and $h$, $\lambda$ are Higgs couplings. The last five terms
represent SUSY-breaking.
Equation (3) has two coupling constants $h$ and $\lambda\,$; we
will study the strong coupling behaviour when {\it h} is very large
and hence for simplicity we  set $\lambda=0$. Also we take
$m_1=m_2=m$ for simlicity.

 It is more convenient to work with the fields defined by
$$\chi_{1,2}={1\over \sqrt 2}\bigl(\Phi_1\pm\Phi_2\bigr)\eqno{(4)}$$
 Now the Higgs potential reduces to
$$\eqalignno{V=\,&h^2\bigl\vert N\bigr\vert^2\bigl(\chi_1^\dagger\chi_1
+\chi_2^{\dagger}\chi_2\bigr)+{1 \over 4}h^2\bigl\vert \chi_1^\dagger\chi_1-
\chi_2^\dagger\chi_2+\chi_2^\dagger\chi_1-\chi_1^\dagger
\chi_2\bigr\vert^2 &\cr
&+\bigl(m^2-m_3^2\bigr)\chi_1^\dagger\chi_1+\bigl(m^2+m_3^2\bigr)
\chi_2^\dagger\chi_2
+m_4^2 N^{\ast}N+m_5^2\bigl(N^2+N^{\ast2}\bigr)&(5)\cr}$$
\vskip .2 cm
\noindent We now assume that the fields $\chi_{1,2}$, where
$$\chi_k\equiv{1\over\sqrt2}\pmatrix{\chi_{kR}^c+i\chi_{kI}^c\cr
\chi_{kR}^0+i\chi_{kI}^0\cr},\quad  (k=1,2)\eqno{(6)}$$
(the   superscripts   c   and  0  denoting  charged  and  neutral
components),
break  the SU(2)$\,\bigotimes\,$U(1) symmetry by assuming a non-zero
vacuum expectation value (VEV)
$$\langle\chi_1\rangle=\pmatrix{0\cr v\cr},\quad \langle\chi_2
\rangle=\pmatrix{0\cr 0\cr}\eqno{(7)}$$
All other fields in (5) are assumed to have zero VEV's. Writing
$N={1 \over 2}(N_1+iN_2)$, there are ten real fields in (5) :
$\chi_{1R}^c\,$, $\, \chi_{1I}^c\,$, $\, \chi_{1R}^0\,$, $\, \chi_{1I}^0\,$,
$\, \chi_{2R}^c\,$, $\, \chi_{2I}^c\,$, $\, \chi_{2R}^0\,$, $\,\chi_{2I}^0\,$,
$\, N_1$ and $N_2$, which we denote by $\eta_i\, (i={\rm 1\, to \,10})$
respectively and their tree level masses by $M_i$.
Even after $\chi_1$ develops a non-zero VEV, the Higgs potential
has a residual symmetry. To see this explicitly we define\thinspace:
$$\vec G\equiv \bigl(\chi_{1I}^c\, , \, -\chi_{1R}^c\, ,\,
  \chi_{1I}^0\bigr) \equiv \bigl(\eta_2,\, -\eta_1,\, \eta_4\bigr)\, ,$$
$$\vec H\equiv \bigl(\chi_{2R}^c\, , \, \chi_{2I}^c\, ,
  \, \chi_{2R}^0 \bigr)\equiv \bigl(\eta_5,\, \eta_6, \, \eta_7 \bigr)$$
$$S_1=\chi_{1R}^0\equiv\eta_3 \, \quad {\rm and} \quad S_2=\chi_{2I}^0
\equiv \eta_8$$
Now shifiting the fields by their VEV's, Higgs potential can
be written as
$$\eqalign{V=&\hskip .2cm {1\over 4}h^2\bigl(N_1^2+N_2^2\bigr)
      \Bigl[{\vec G}^2+\bigl(S_1+v\bigr)^2+{\vec H}^2+S_2^2\Bigr]\cr
   &+{1 \over 16}h^2\Bigl[{\vec G}^2+\bigl(S_1+v\bigr)^2-\vec H^2
   -S_2^2\Bigr]^2+{1 \over 4}h^2\Bigl[\vec G.\vec H -\bigl(S_1+v\bigr)
   S_2\Bigr]^2\cr
   &+{1 \over 2}\bigl(m^2-m_3^2\bigr)\Bigl[\vec G^2+\bigl(S_1+v\bigr)
   ^2\Bigr]
    +{1 \over 2}\bigl(m^2+m_3^2\bigr)\bigl(\vec H^2+S_2^2\bigr)\cr
   &+{1 \over 2}\bigl( m_4^2+2m_5^2\bigr)N_1^2
    +{1\over 2}\bigl(m_4^2-2m_5^2\bigr)N_2^2 \cr} $$
\vskip .2cm
\noindent Under a SU(2) rotation wherein $\vec G$ and $\vec H$
are triplets, and $\bigl(S_1,\, S_2,\, N_1,\, N_2\bigr)$ all
singlets, V is invariant. We expect then $M_1=M_2=M_4$ and
$M_5=M_6=M_7$ on account of this symmetry. We note that this
symmetry is present in a two-doublet model for the special choice of
our parameters. Thus in the two-doublet model discussed in section 2
of reference [8], setting $v_1=v_2=\xi=0$
and $\lambda_1=\lambda_4$ makes $H_1^0$, $H_2^0$ and $H_4^0$
degenerate. We will build in this residual symmetry in our
variational approach.

 In terms of ten real fields the Higgs potential
 $$\eqalignno{V=&\hskip .2 cm \bigl(m^2-m_3^2\bigr)
 \thinspace v^2 \thinspace +{1 \over 4 }h^2 v^4 \thinspace
+{1 \over 2}\bigl(m^2-m_3^2+{1 \over 2}h^2 v^2\bigr)
\thinspace \bigl(\eta_1^2+\eta_2^2 +\eta_4^2\bigr)&\cr
&+{1 \over 2}\bigl(m^2-m_3^2+{3 \over 2}h^2 v^2\bigr)\eta_3^2
+{1 \over 2}\bigl(m^2+m_3^2-{1\over 2}h^2 v^2\bigr)\bigl(\eta_5^2
+\eta_6^2+\eta_7^2\bigr)&\cr
&+{1 \over 2}\bigl(m^2+m_3^2+{1\over 2}h^2 v^2\bigr)\eta_8^2
+{1 \over 2}\bigl(m_4^2+2m_5^2+h^2 v^2\bigr)\eta_9^2&\cr
&+{1 \over 2}\bigl(m_4^2-2m_5^2+h^2 v^2\bigr)\eta_{10}^2
 +{1 \over 16}h^2\sum_{i=1}^8\eta_i^4
+{1 \over 4}h^2\bigl(\eta_9^2+\eta_{10}^2\bigr)\sum_{i=1}^8\eta_i^2&\cr
 &+{1 \over 8}h^2\Bigl[\eta_1^2\bigl(\eta_2^2+\eta_3^2+\eta_4^2
  -\eta_5^2+\eta_6^2-\eta_7^2-\eta_8^2\bigr)&\cr
 & +\eta_2^2\bigl(\eta_3^2+\eta_4^2+\eta_5^2-\eta_6^2
   -\eta_7^2-\eta_8^2\bigr) +\eta_3^2\bigl(\eta_4^2-\eta_5^2
   -\eta_6^2-\eta_7^2+\eta_8^2\bigr)&\cr
 &+\eta_4^2\bigl(-\eta_5^2-\eta_6^2+\eta_7^2-\eta_8^2\bigr)
 +\eta_5^2\bigl(\eta_6^2+\eta_7^2+\eta_8^2\bigr)&\cr
 &+\eta_6^2\bigl(\eta_7^2+\eta_8^2\bigr)+\eta_7^2\eta_8^2
 \Bigr]+V_{linear}+V_{cubic} &(8) \cr}$$
\vskip .2 cm
\noindent where $V_{linear}$ and $V_{cubic}$ respectively represent the terms
linear and cubic in fields. The tree level minima condition is
equivalent to equating the term linear in $\eta_3$
$\bigl(${\it i.e.}$\,\chi_{1R}^0\bigr)$ in (8) to zero,
$$m^2-m_3^2+{1 \over 2}h^2v^2=0\eqno{(9)}$$
Inspection of the qudratic terms in (8) together with (9)
immediately tells us that
$$M_1=M_2=M_4=0\, ,\eqno{(10)}$$
indicating that $\eta_1$, $\eta_2$ and $\eta_4$ are the goldstones.
Furthermore, we have for the charged Higgs triplet
$$M_5^2=M_6^2=M_7^2=m^2+m_3^2-{1 \over 2}h^2 v^2 \eqno{(11)}$$
where the degeneracy is as expected. Lastly,
$$ M_3^2=m^2-m_3^2+{3 \over 2} h^2 v^2,\eqno{(12)}$$
$$ M_8^2=m^2+m_3^2+{1 \over 2} h^2 v^2,\eqno{(13)}$$
$$ M_9^2=m_4^2+2m_5^2+h^2 v^2 \eqno{(14a)}$$
and
$$M_{10}^2=m_4^2-2m_5^2 +h^2 v^2\eqno{(14b)}$$
are the masses of neutral Higgs. For simplicity we also take
$m_5=0$ so that
$$M_9^2=M_{10}^2 \eqno{(15)}$$

\beginsection {3. Gaussian Trial Wave-Functional}

  \indent In order to obtain information on masses beyond the tree level,
 we follow a variational method with a Gaussian trial wave-
functional. Most generally, this wave functional would be the
vacuum state of a set of free fields of masses $\Omega_1,\ldots,
\Omega_{10}$ with the $\Omega's$ (and {\it v} in equation (7))
representing the variational parameters. However, taking a variational
ground state that respects the residual symmetry stated in the
last section, we set $\Omega_1=\Omega_2=\Omega_4$
and  $\Omega_5 = \Omega_6 = \Omega_7$. Further imposing the symmetry
in (15) we also put $\Omega_9 = \Omega_{10} = \Omega_N$. We thus have five
 independent masses $\Omega_1$, $\Omega_3$, $\Omega_5$,
$\Omega_8$ and $\Omega_N$, and of course $v$, as variational
parameters.

Following standard techniques [9], the expectation value of
Hamiltonian density ${\cal H}$ in our trial vacuum state wave functional
is
$$ \eqalignno{V_G=&\thinspace\bigl(m^2-m^2_3\bigr) v^2
     + {1 \over 4}h^2 v^4 +2\biggl[{1 \over 2}\bigl(M_N^2
     -\Omega_N^2\bigr)I_0(\Omega_N^2)+I_1(\Omega_N^2)\biggr] &\cr
    &+\sum_{i=1}^8\biggl[{1 \over 2}\bigl(M_i^2-\Omega_i^2\bigr)
I_0(\Omega_i^2)+I_1(\Omega_i^2)\biggr]+{1 \over
2}h^2I_0\bigl(\Omega_N^2\bigr)\sum_{i=1}^8I_0(\Omega_i^2) &\cr
&+{3 \over 16}h^2\biggl[5I_0^2(\Omega_1^2)+
    I_0^2(\Omega_3^2)+5I_0^2(\Omega_5^2)+I_0^2(\Omega_8^2)\biggr] &\cr
   &+{1 \over 8}h^2\biggl[3I_0(\Omega_1^2)\Bigl(I_0(\Omega_3^2)
     - I_0(\Omega_5^2)-I_0(\Omega_8^2)\Bigr)&\cr
      &+3I_0(\Omega_5^2)\Bigl(I_0(\Omega_8^2)-I_0(\Omega_3^2)\Bigr)
      +I_0(\Omega_3^2)I_0(\Omega_8^2)\biggr] &(16)\cr}$$
 Here
 $$ I_1(\Omega)={1 \over (2\pi)^4}\int {d^4k_E\;
 {\rm ln}\bigl(k_E^2+\Omega^2\bigr)} \quad +{\rm constant}\eqno{(17a)}$$
 and
$$  I_0(\Omega)={1 \over (2\pi)^4}\int {d^4k_E{1\over\bigl
  (k_E^2+\Omega^2\bigr)}}\eqno{(17b)}$$
\vskip .2 cm
\noindent  Differentiating (16) {\it w.r.t.} $\Omega_N^2$, $\Omega_1^2$,
 $\Omega_3^2$, $\Omega_5^2$ and $\Omega_8^2$ gives us five
 mass equations\thinspace:
$$\eqalignno{\Omega_N^2&=M_N^2+{1\over 2}h^2\Bigl[3I_0(\Omega_1^2)
    +3I_0(\Omega_5^2)+I_0(\Omega_3^2)+I_0(\Omega_8^2)\Bigr]&(18a)\cr
\Omega_1^2&=M_1^2+{1\over 4}h^2 \Bigl[4I_0(\Omega_N^2)
+5I_0(\Omega_1^2)+I_0(\Omega_3^2)
  -I_0(\Omega_5^2) - I_0 ( \Omega_8^2)\Bigr]&(18b)\cr
\Omega_3^2&=M_3^2+{1 \over 4}h^2\Bigl[4I_0(\Omega_N^2)
+3I_0(\Omega_1^2)+3I_0(\Omega_3^2)-3I_0(\Omega_5^2)
+I_0(\Omega_8^2)\Bigr] &(18c)\cr
\Omega_5^2&=M_5^2+{1 \over 4}h^2\Bigl[4I_0(\Omega_N^2)
-I_0(\Omega_1^2)-I_0(\Omega_3^2)+5I_0(\Omega_5^2)
+I_0(\Omega_8^2)\Bigr]&(18d)\cr
\Omega_8^2&=M_8^2+{1\over 4}h^2\Bigl[4I_0(\Omega_N^2)
-3I_0(\Omega_1^2)+I_0(\Omega_3^2)+3I_0(\Omega_5^2)
+   3   I_0 (\Omega_8^2)\Bigr]&(18e)\cr}$$
 \vskip .2 cm
\noindent Using equations (18), $V_G$  reduces to
 $$\eqalignno{V_G=&\thinspace ( m^2 - m^2_3) v^2 + {1\over 4}h^2 v^4 & \cr
  &+\biggl[2I_1
  \bigl(\Omega_N^2\bigr)+3I_1\bigl( \Omega_1^2\bigr) + I_1 \bigl(
  \Omega_3^2  \bigr)  + 3 I_1 \bigl( \Omega_5^2 \bigr) + I_1 \bigl(
  \Omega_8^2 \bigr)\biggr]&\cr
  & -{3 \over 16}h^2 \biggl[ 5I_0^2(\Omega_1^2)
 +I_0^2(\Omega_3^2)+5I_0^2(\Omega_5^2)+I_0^2(\Omega_8^2)\biggr]&\cr
  &-{1\over2}h^2I_0(\Omega_N^2)\biggl[3I_0(\Omega_1^2)+I_0(\Omega_3^2)
 +3I_0(\Omega_5^2)+I_0(\Omega_8^2)\biggr]&\cr
 &-{1\over8}h^2\biggl[3I_0(\Omega_1^2)\Bigl(I_0(\Omega_3^2)
 -I_0(\Omega_5^2)-I_0(\Omega_8^2)\Bigr)&\cr
  &\hskip 1 cm +3I_0(\Omega_5^2)\Bigl(I_0(\Omega_8^2)
 -I_0(\Omega_3^2)\Bigr)
 +I_0(\Omega_3^2)I_0(\Omega_8^2)\biggr]&(19)\cr}  $$
\vskip .2 cm
\noindent Here $\Omega_i$'s are to be understood as depending on
$v$ through (18). Differentiating $V_G$
{\it w.r.t.} $v^2$, we get
 $$ \eqalignno {{dV_G \over dv^2} =&\thinspace m^2 -m_3^2
+ {1 \over 2} h^2 v^2 + {1\over 4}h^2 \Bigl[4I_0(\Omega_N^2)
+ 3 I_0 (\Omega_1^2) &\cr
&\hskip 4cm + 3I_0 (\Omega_3^2)- 3I_0 (\Omega_5^2) + I_0(\Omega_8^2)\Bigr]
&(20)\cr}$$
Setting $d V_G /dv^2$ to zero and using (18), we get
$$ \Omega_3^2 = \thinspace h^2 v^2 \eqno {(21)}$$
It is clear from (20), that increasing $h$ would increase
$\Omega_3^2$
without limits. However, the limit would be set by demanding that
stationary solution (20) be stable, {\it i.e.}
the stability matrix $\Bigl(\partial^2 V_G\Bigr)$ be positive definite.
Stability condition is obtained by considering $\Bigl(\partial^2V_G\Bigr)$
to be a function of six variables --- five mass parameters
$\bigl(\Omega_1^2$, $\Omega_3^2$, $\Omega_5^2$, $\Omega_8^2$ and
$\Omega_N^2\bigr)$ and $v^2$, and demanding all its eigenvalues
to be positive. As in the case
of standard model [4], we expect that as $h$ increases, this
condition would no longer be satisfied beyond a certain
maximum value of $h$.

 Our  query  regarding  triviality  bounds  does  not involve the
complete   numerical   solution.  Of  five  independent  masses
$\Omega_1, \Omega_3, \Omega_5, \Omega_8$ and $\Omega_N$,
$\Omega_1$ is  the mass of the Goldstone bosons. This in an exact
calculation  is  expected  to  be zero but in variational methods
(see reference [9] for elaboration ),      we can get a small but
non-vanishing mass. Of the remining four Higgs masses, we wish to
find  out whether they can be made arbitrarily heavy relative  to
${\it v}^2$ (or $M_W^2$). The only condition we would impose   is
the same as laid  down  by  Hasenfratz and Nager, namely that
for a cut off theory to make any  physical  sense,  each  one  of
the masses $\Omega_i$ must not be close to or greater than the
cut-off\thinspace; we  put  an  upper limit of $\Lambda_c/5$ for
definiteness, for all $\Omega_i$'s (See reference [2]).

 Our  task  is then to set $\Omega_i$'s at their maximum
possible values and determine the value of $h$ for which the
eigen values of stability matrix go from positive to negative.
Since there are only three mass input parameters $m$, $m_3$ and
$ m_4$, all the five
$\Omega_i$'s cannot be assigned arbitrary values by suitably
choosing $m$'s. Furthermore, only $\Omega_N^2$ involves $m_4^2$,
so that we can set $\Omega_N$ as the highest acceptable mass
namely $\Lambda_c /5$ right away. Of the remaining four masses,
we immediatly have the sum rule,
$$\eqalignno{\Omega_8^2-\Omega_5^2&=\Omega_3^2-\Omega_1^2&\cr
   &=h^2v^2+{h^2 \over 2}\Bigl[I_0(\Omega_3^2)-I_0(\Omega
   _1^2)+I_0(\Omega_8^2)-I_0(\Omega_5^2)\Bigr]&(22)\cr}$$
Also since $I_0(\Omega^2)$ is a decreasing function of $\Omega^2$,
we get from (18) and (22)
$$\Omega_8^2>\Omega_5^2 \quad {\rm and} \quad \Omega_3^2>\Omega_1^2$$
The first possibility is to set
$$\Omega_8=\,\Omega_3=\,\Lambda_c/5 \, ;$$
$\Omega_1$ is the Goldstone and $\Omega_5 $ is then the lightest
Higgs mass. However in this case $\Omega_5 = \Omega_1$, and the
maximum value of $\Omega_5$ turns out to be lower than in the next
sequence of masses.

 Next we assume $\Omega_8 = \Lambda_c/5$ and the sequence
$$\Omega_8 > \Omega_5 >\Omega_3 >\Omega_1$$
In this $\Omega_3$ is the lightest Higgs. From (22), pushing
$\Omega_5 $ towards $\Omega_8$ will make $\Omega_3$ go towards
$\Omega_1$, {\it i.e.} will make $\Omega_3$ lighter. We expect
then some kind of optimal situation to arise if
$\Omega_5=\Omega_3$.
With the values of $\Omega_3$ (obtained from (18)) corresponding
for various $h$, we can calculate the eigen values of stability
matrix. As expected one eigen value crosses over from positive
to negative at a value of Higgs coupling,
$$h\equiv h_{max}=4.52 \eqno{(23)}$$
Using this value, then the upper bound on mass of the lightest
Higgs $M_{LH}=\Omega_3=\Omega_5$ is
$${M_{LH} \bigl\vert_{max} \over M_W} =\, 10.1\eqno{(24)}$$
 The Goldstone mass $\Omega_1$  for this choice, as we stated before,
 is not zero but smaller than other masses in the spectrum.

\beginsection {4. Conclusion}

\indent We have shown that in the next to minimal version of the supersymmetric
model, the lightest Higgs particle has an upper bound which is
$\sim 10 M_W$. This is of the same order as the bounds in the
standard model. We have not  attempted to determine the absolute
bound taking into account the full range of variation of parameters
in the NMSSM (including SUSY breaking parameters). This is because
our main aim was  to show that in the non minimal version of
supersymmetric model, one is not constrained by the rather strict
limits on the Higgs mass that one obtains in the minimal model.
Also limits on the Higgs mass above 1 TeV are of little interest.
There is no possibility in any near future to detect any signals
for such a heavy Higgs. Moreover Higgs particle with masses above 1 TeV
with widths comparable to masses will make the mass
parameter rather meaningless from an experimental point of view.
What is more relevant is to note that in  a supersymmetric theory,
the Higgs mass bounds has the same features
 as the regular standard model.

\beginsection {References}

{\parindent 15 pt
\item{1.} For a summary, see K. Huang, MIT preprint, CTP \# 1631,1988.
\item{2.} P. Hasenfratz and J. Nager, Zeit. Phys. {\bf C37}, 477 (1988).
\item{3.} M. Luscher and P. Weisz, Nucl. Phys. {\bf B290}, 25 (1987),
{\bf B295}, 65 (1988).
\item{4.} S.R. Choudhury and Mamta, Delhi preprint (submitted for
publication).
\item{5.} See {\it e.g.} J.F. Gunion {\it et. al.}, 'The Higgs Hunter's
Guide' (Addison-Wesley, 1990).
\item{6.} M. Quiros, CERN preprint, CERN-TH. 7507/94 (November 1994).
\item{7.} Y.Y. Wu, Phys. Rev. {\bf D51}, 5276 (1995)
\item{8.}J.F. Gunion and H.E. Haber, Nucl. Phys. {\bf B272}, 1 (1986)
\item{9.} P.M. Stevenson, B. ${\rm All\grave es}$ and R. Tarrach, Phys.
Rev. {\bf D35}, 2407 (1987).
}
\bye